\documentclass[a4paper]{article}

\usepackage{INTERSPEECH2021}

\title{Zero-Shot Text-to-Speech for Text-Based Insertion in Audio Narration}
\name{Chuanxin Tang$^1$, Chong Luo$^1$, Zhiyuan Zhao$^1$, Dacheng Yin$^2$, Yucheng Zhao$^2$, Wenjun Zeng$^1$}
\address{
  $^1$Microsoft Research Asia, Beijing, China\\
  $^2$University of Science and Technology of China, Hefei, China}
\email{\{chutan, cluo, zhiyzh, wezeng\}@microsoft.com, \{lnc, ydc\}@mail.ustc.edu.cn}

\begin{document}

\maketitle
\begin{abstract}

  Given a piece of speech and its transcript text, text-based speech editing aims to generate speech that can be seamlessly inserted into the given speech by editing the transcript.
  Existing methods adopt a two-stage approach: synthesize the input text using a generic text-to-speech (TTS) engine and then transform the voice to the desired voice using voice conversion (VC). A major problem of this framework is that VC is a challenging problem which usually needs a moderate amount of parallel training data to work satisfactorily. 
  In this paper, we propose a one-stage context-aware framework to generate natural and coherent target speech without any training data of the target speaker. In particular, we manage to perform accurate zero-shot duration prediction for the inserted text. The predicted duration is used to regulate both text embedding and speech embedding. Then, based on the aligned cross-modality input, we directly generate the mel-spectrogram of the edited speech with a transformer-based decoder. Subjective listening tests show that despite the lack of training data for the speaker, our method has achieved satisfactory results. It outperforms a recent zero-shot TTS engine by a large margin.

\end{abstract}

\section{Introduction}


In the new norm of work and entertainment under COVID-19, people are increasingly using audio and video recordings. In 2020, most of the international conferences, if not cancelled, were going online. In these online conferences, presenters are often required to record their speech beforehand to avoid possible network interruptions during a live broadcast. Podcast usage is also seeing a rapid growth, by 42\% from March, 2020 to April, 2020, according to Voxnest \cite{voxnest.org}. 

One of the main pain points when people record audios/videos is that it is hard to make changes when a mistake is made, even if it is just a single misspoken word. Recently, several user-friendly audio/video editing tools \cite{jin2017voco,fried2019text,discript.org} are proposed, allowing users to make small changes to audio narration or even video frames by editing transcript. Such systems support the deletion, cutting/copying, cutting/pasting, insertion, and replacement of a word or a short phrase in the transcript with corresponding changes automatically made in the audio or video. While deletion, cutting/copying, and cutting/pasting only requires cross-modality alignment and a stitching technique, insertion and replacement involves the generation of unseen words and therefore is much more difficult to handle. 


In this paper, we shall address the zero-shot text-based speech insertion problem. Given a piece of speech and its transcript with the inserted word, we aim to automatically synthesize the inserted word in a voice that sounds seamless in context – as if it were uttered by the same person in the same recording. We ambitiously target at the case where no training corpus of the target speaker is available except for the context speech.
A classic solution of this problem is to use a TTS synthesizer to say the word in a generic voice, and then use voice conversion (VC) to convert it into a voice that matches the narration \cite{jin2017voco}. VC is achieved by unit selection in the pioneering work CUTE \cite{jin2016cute} and a later system VoCo \cite{jin2017voco}. 
The source voice is used as a query to search for snippets of the target voice that sound similar to the query as much as possible, while also matching at the stitch boundaries through minimizing the predefined loss function. A main drawback of this unit selection based VC is that it needs 20 to 40 minutes of target voice \cite{jin2017voco} to achieve a good performance. 
Some other works \cite{su2020acoustic,morrison2021context} explore attributes-based voice conversion. They use a predefined method to extract certain types of attributes of the target speech. These methods can improve the speech quality to a certain extent, but an inherent limitation still exists. If the gap between the TTS-synthesized voice and the target voice is large, or attributes other than extracted ones are different, there is no quality guarantee. Again, these data-driven VC methods require a moderate amount of parallel speech corpus (source and target voice with the same content). 

\begin{figure}[t]
  \centering
  \includegraphics[width=\linewidth]{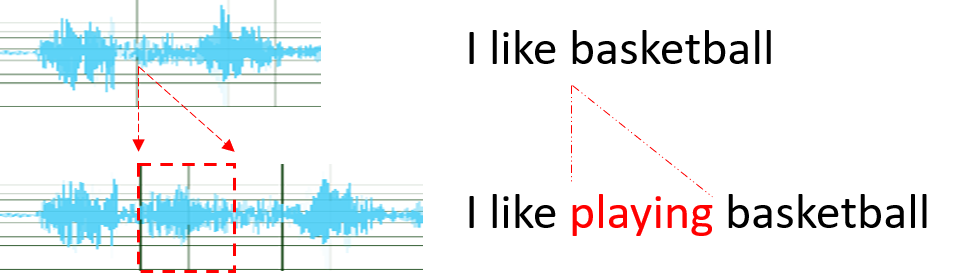}
  \caption{Text-based speech editing. Through editing text, synthesize the speech of the inserted word (red) that sounds seamless in context (blue) – as if it were uttered by the same person in the same recording. }
  \label{fig:task_desc}
\end{figure}


Another line of research that can potentially be applied in the scenario of our interest is zero-shot TTS \cite{jia2018transfer}. Although the mainstream TTS research \cite{shen2018natural,li2019neural,ren2019fastspeech,ren2020fastspeech} focuses on synthesizing high-quality speech using certain timbre with sufficient training corpus, zero-shot TTS \cite{jia2018transfer} aims to synthesize any speaker's speech without training on that speaker's corpus. The main idea is to use a pre-trained speaker verification architecture to extract the target speaker embedding from the reference speech, and then synthesize speech for new content. However, only the global feature of the speaker, such as timbre, is considered in generating the target word. The context given is not utilized for smoothness or naturalness.


Speech information can be roughly decomposed into three components \cite{qian2020unsupervised}: language content, timbre, and prosody. In order to synthesize speech that can be seamlessly inserted into the reference speech, these three components of the speech should be jointly considered. Unfortunately, none of the existing methods have done so. We point out that transcript text provides content information, reference speech provides timbre information, and the context surrounding the inserted words provides prosody information. It is based on this observation that we design a one-stage context-aware framework for zero-shot text-based speech synthesis. Our framework has three important features. First, it is a one-stage framework that directly synthesizes the target speech by simultaneously considering all the input information. Second, it is context-aware by applying transformer to the aligned multi-modality hidden sequences. Third, as far as we know, ours is the first zero-shot approach to achieve high-quality results in this scenario. 


\begin{figure}[t]
  \centering
  \includegraphics[width=\linewidth]{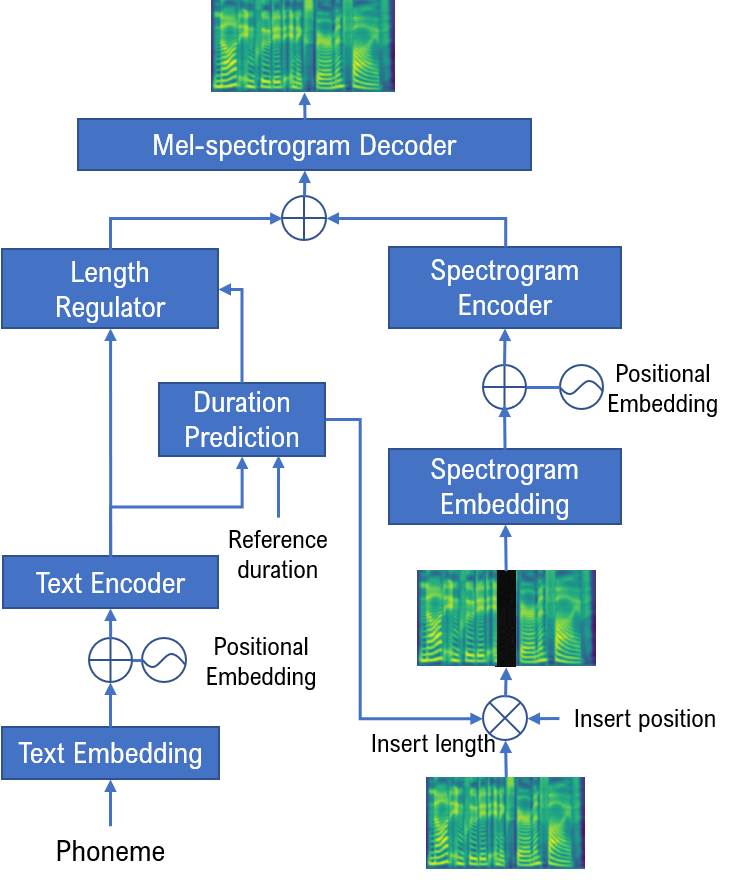}
  \caption{Proposed one-stage context-aware text-based speech editing framework. We align the phoneme sequence and the mel-spectrogram sequence before feeding the sum of the two to the mel-spectrogram decoder to directly generate the desired speech. $\otimes$ represents zero padding operation.}
  \label{fig:framework}
\end{figure}

\section{Proposed Methods}
\subsection{Overview}
Given a piece of speech, its transcript text, and the edited transcript with an inserted word or short phrase, we aim to generate a natural and coherent edited speech for the edited transcript, without any additional prior information or training data about the speaker. There are at least three requirements to ensure naturalness and coherence of the generated speech: 1) The inserted speech should match the inserted word in the edited transcript and should have sufficient intelligibility; 2) The timbre of inserted speech should be the same as that of the speaker; 3) The prosody of the generated speech, such as tone and rhythm, should be coherent with its surrounding speech.
A major challenge in solving this problem is to meet these requirements without any training data of the given speaker. This is a constrained zero-shot speech synthesis problem. 

We propose a novel one-stage context-aware framework to solve this problem. There are two prominent features in our solution. First, different from conventional two-stage solutions which involve an intermediate speech generated by a generic TTS engine, our solution directly produces the mel-spectrogram of the target speech from the inserted phonemes and the mel-spectrogram of the original speech. No generic TTS is required. Second, our solution is context-aware. When the same word is inserted into different parts of a speech, the synthesized speech could be different in tone and prosody. Similarly, when the same word is inserted into the speech of different speakers, the synthesized speech will be different in timbre as well. 


This work makes two additional technical contributions. First, inspired by a generic TTS engine named FastSpeech2 \cite{ren2020fastspeech} which benefits greatly from duration estimation, we introduce a duration estimation module. However, conventional TTS engine uses a large amount of training data of a specific speaker to learn the phoneme duration to be spoken by this speaker. In our scenario, the speaker is unknown during training, making it a challenging zero-shot prediction problem. 
Second, we use transformers in multiple modules in the framework to explore the global correlation and make full use of the provided speech and transcript. We believe that it is because of transformer's superior attention ability that the challenging zero-shot TTS problem can be solved satisfactorily.


Our framework is demonstrated in Figure ~\ref{fig:framework}. It can be decomposed into two phases. In the first multi-modal alignment phase, the duration of the inserted text is estimated and then the phoneme embedding and the speech embedding are matched and aligned. In the second decoding phase, the cross decoder module, built upon transformers, fully explores the local and global correlations in the input and directly generates the desired mel-spectrograms. Finally, for simplicity, the Griffin-Lim \cite{griffin1984signal} algorithm is applied as vocoder to synthesize waveforms.

\subsection{Multi-modal alignment phase}

This phase handles two streams: phoneme stream and mel-spectrogram stream. For the phoneme stream, the input phoneme (converted from the entire transcript including the inserted words) is first processed by a text embedding block. This block first embeds the input one-hot vector into a trainable embedding of 512 dimensions and then uses a one-dimensional 3-layer CNN with ReLU activation to produce an embedding which contains context information in the input phoneme sequence. Batch normalization, ReLU activations and dropout are applied after every CNN layer. An extra linear layer is added to project the hidden phoneme sequence to the embedding of 256 dimensions.

Then, a scaled positional encoding \cite{li2019neural} is added to the text embedding before it is fed into the text encoder. The text encoder is a 2-layer transformer encoder with multi-head self-attention and position-wise fully connected (FC) feed-forward network \cite{vaswani2017attention}. The output of the encoder is the hidden phoneme sequence. It is first sent to the duration prediction module and then, with the predicted duration, it passes through a length regulator to obtain 
the extended hidden phoneme sequence.

Here, phoneme duration is defined as the length of the phoneme in the mel-spectrogram, as in \cite{ren2019fastspeech}. In our framework, we perform zero-shot prediction of phoneme length in the duration prediction module. In addition to the encoded phoneme sequence, the reference duration is also provided to the predictor. In particular, duration of existing phonemes is obtained from Montreal forced alignment (MFA) tools \cite{mcauliffe2017montreal} and that of the inserted phonemes are set to zero. 
The duration prediction module consists of one transformer encoder layer and 2 FC layers. ReLU activations are applied after every FC layer except the last one. 
The predicted duration is used to regulate both phoneme sequence and the mel-spectrogram sequence. 

For the mel-spectrogram stream, the input is first extended with the information of the edited position and the predicted duration. Specifically, we apply zero padding to the mel-spectrogram corresponding to the edited region. As such, the extended mel-spectrogram has the same length as the extended hidden phoneme sequence. 
The extended mel-spectrogram is then consumed by the spectrogram embedding module which is a neural network composed of two FC layers (each has 256 hidden units) with ReLU activation followed by a scaled positional encoding as applied in \cite{li2019neural}. Finally, the extended mel-spectrogram with the positional encoding information is processed by the spectrogram encoder to obtain the extended hidden mel-spectrogram sequence. The spectrogram encoder has the same network structure as the text encoder. 

It can be noticed that transformers are extensively used in our framework. The idea behind this design is that transformer is good at capturing long-range dependency for a sequence. 
We apply zero padding to the input spectrogram before positional embedding. With transformers, 
the model can gradually "inpaint" the hole at the edited regions. Actually, we have also tried to 
omit this step, but the results are far from satisfactory. 


\subsection{Decoding phase}

We perform position-wise addition for the aligned phoneme and mel-spectrogram sequences. Then, we adopt a non-autoregressive decoder to predict the synthesized mel-spectrogram from the combined hidden sequence. The reasons why we use addition operation are two-fold. First, we want the input and the output of the decoder to have the identical length, so that a non-autoregressive decoder can be applied. It is known that we have to use an autoregressive decoder if there is length discrepancy between the input and the output. Second, the model can have explicit matching information between the phoneme sequence and the mel-spectrogram sequence. 

The decoder module is implemented by a 5-layer transformer encoder followed by a linear layer. We use L2 loss to measure the distance between the synthesized and the ground-truth mel-spectrogram. L1 loss is applied to calculate the distance between the predicted and the ground-truth phoneme duration. The weights for the two losses are 1 and 0.01, respectively. It is worth noting that loss is applied on the entire sentence, not just the edited region.

\section{Experiments}
\subsection{Experiment setup}
\subsubsection{Dataset}

We conduct experiments on LibriTTS dataset \cite{zen2019libritts}, which contains both English audio clips and the corresponding text transcripts of 2456 speakers. The total audio length is about 585 hours. All of our models are trained on the \emph{train-clean-360} subset which contains 430 female and 474 male speakers, with the total audio length of about 191 hours. All speech recordings are downsampled to 24 kHz sampling rate.
In order to solve mispronunciation problems \cite{shen2018natural}, we use the Montreal forced alignment (MFA) \cite{mcauliffe2017montreal} tool to convert text to phoneme and acquire alignment information between speech and phoneme.

\subsubsection{Implementation details}

In text embedding block, the kernel size of CNN layer is 5 and dropout rate is set to 0.2. The hidden size of the self-attention are set to 256 and the number of attention heads is set to 4.

We use four Nvidia Tesla V100 to train our model. The ADAM \cite{kingma2014adam} optimizer with default parameters is applied for 100 epochs. Learning rate is 1e-3 and batch size is 32. Missing words are randomly chosen during training and testing. 

We use mel-spectrograms with 80 bins using librosa mel filter defaults. Mel-spectrograms are computed through a short time Fourier transform (STFT) using a 50ms frame size, 12.5 ms frame hop, and a Hann window function.

We use the Griffin-Lim algorithm \cite{griffin1984signal} as vocoder to synthesize waveform from the predicted mel-spectrogram. A CBHG module \cite{wang2017tacotron} is applied to predict spectral magnitude sampled on a linear-frequency scale. It is trained separately on \emph{train-clean-360} subset of LibriTTS. The hyperparameters of Griffin-Lim and CBHG are the same as in \cite{wang2017tacotron}. We emphasize that our choice of Griffin-Lim is for simplicity. Developing a fast and high-quality trainable spectrogram to waveform converter is an ongoing work.

\subsubsection{Evaluation methods}
We primarily rely on Mean Opinion Score (MOS) evaluations based on subjective listening tests with rating scores from 1 to 5. We evaluate synthesized speech along two dimensions: its naturalness and similarity to real speech from the target speaker. We randomly choose four speakers (two males and two females) from LibriTTS test-clean set. For each speaker, ten sentences are randomly selected. The chosen speakers are not included in the training set. The following experiments are based on these sentences. We use the method in \cite{jia2018transfer} as baseline for comparison. The target sentence is fed as reference speech for \cite{jia2018transfer} to extract speaker features. However, since the baseline method is not developed for word insertion, it does not make use of the insertion position and therefore does not provide smoothness guarantee. 

\subsection{Speech naturalness}
We conducted three experiments to evaluate the speech naturalness of our methods. The first one is an identification test where subjects will tell whether they think a narration contains an inserted synthetic word. The second one is a Mean Opinion Score (MOS) test that asks subjects to rate the quality of the inserted synthetic words. The third one is an objective test to evaluate the accuracy of phoneme duration prediction.

For each sentence selected, we maintain three versions including the original recording. To create the other two altered versions, we randomly choose and remove one word, and synthesize it using our method or the voice clone method as described in \cite{jia2018transfer}. We ensure that the selected word has more than one phoneme. Then we insert the synthesized word to its original location to create the altered narrations. 

\subsubsection{Speech naturalness identification test}
In the identification test, a subject is presented one narration at a time and is informed that there may or may not be one synthesized word. The task is to identify if the narration is original or edited. We use original rate (the number of sentences that are considered as original recordings divided by the total number of sentences) to evaluate the performance of the model. A higher original rate indicates that more synthesized sentences are considered to be original. 

\begin{table}[th]
  \caption{Speech naturalness identification test}
  \label{tab:identification test}
  \centering
  \begin{tabular}{l  r }
    \toprule
    \textbf{Method} & \textbf{Original Rate} \\
    \midrule
    Ground truth                        & 94.75\%              \\
    \hline
    Baseline\cite{jia2018transfer}      & 17.75\%              \\
    Ours                                & \textbf{47.25}\%     \\
    \bottomrule
  \end{tabular}
\end{table}

Table \ref{tab:identification test} shows the results. The original recordings receive an original rate of 94.75\%, which can be treated as an upper bound of this task. The fact that original recordings cannot achieve a 100\% original rate also shows a high standard of the subjects. With this high standard, the baseline method only achieves 17.75\% original rate. Our method, in contrast, achieves a much higher score of 47.75\%, which demonstrate the advantage of our method to fully explore the local and global correlations to synthesize the target speech. 

\subsubsection{Speech naturalness MOS test}
In the speech naturalness MOS test, subjects are additionally asked to rate the quality of the sentence on a scale from 1 to 5 (1: very annoying, 2: annoying, 3: slightly annoying, 4: perceptible but not annoying, 5: almost real), no matter they think it is an original recording or an edited narration. Each sentence is rated by ten people and we use the average score as the final score. 

\begin{table}[th]
  \caption{Speech naturalness Mean Opinion Score (MOS) test}
  \label{tab:MOS test}
  \centering
  \begin{tabular}{l  r }
    \toprule
    \textbf{Method} & \textbf{MOS} \\
    \midrule
    Ground truth      & 4.57            \\
    \hline
    Baseline\cite{jia2018transfer}      &  2.66           \\
    Ours                                &  \textbf{3.86}  \\
    \bottomrule
  \end{tabular}
\end{table}

Table \ref{tab:MOS test} shows the results. We can see that the performance of our method is significantly better than the baseline method. The reason is obvious: we take advantage of the context while the baseline does not. Compared to the original sentence, our method still has room for improvement. It should be noted that Griffin-Lim is a simple vocoder and better results can be expected if a higher quality vocoder is used. 

\subsubsection{Duration prediction}
Zero-shot phoneme duration prediction is an important technical contribution of our work. This section evaluates the quality of our duration prediction and demonstrates its importance in creating a natural speech through the comparison with the baseline method. Two objective metrics, namely phoneme-level error and word-level error, are used to evaluate the performance. The phoneme error is defined as the mean absolute difference between the predicted duration and the ground truth duration of a phoneme. 
Each word consists of a few phonemes, and the duration of the word is computed by summing up the duration of all the phonemes. For baseline method, they do not have a duration prediction step, so we only calculate its word-level error based on the duration of synthesized words.

Table~\ref{tab:duration prediction} shows the results. The average word-level error of our method is 5.04 frames while that of baseline is 46.56 frames. This clearly verifies the effectiveness of our zero-shot duration prediction module. With an more accurate duration, more high-quality target speech can be obtained. It partially explains why our method achieves significantly better subjective scores than the baseline method in the previous two evaluations. 
Besides, the average phoneme-level difference of our method is 1.76 frames. Considering the average phoneme duration is 6.28 frames and word duration is 24.05 frames, it still has room for improvement. We will continue to explore speaker-dependent zero-shot duration prediction problem in our future work.

\begin{table}[th]
  \caption{Duration prediction evaluation}
  \label{tab:duration prediction}
  \centering
  \begin{tabular}{l  c  c}
        \toprule
        \textbf{Method} & \textbf{Phoneme-level error} & \textbf{Word-level error} \\
        \midrule 
        Baseline\cite{jia2018transfer}  & - & 46.56\\
        Ours & \textbf{1.76} & \textbf{5.04}\\
        \bottomrule
    \end{tabular}
\end{table}

\subsection{Speaker similarity}
It is hard to evaluate speaker similarity based on a sentence with only one word altered. Therefore, we design a procedure to generate an utterance with all synthesized words. In particular, for each word in a sentence, we remove it and synthesize it using the rest of the sentence as input. We repeat the process multiple times and then concatenate the synthesized words according to their original order in the sentence. For the baseline method \cite{jia2018transfer}, generating a synthesized sentence is quite straightforward, as it was designed to be used in this way. 
During evaluation, 
we pair each synthesized utterance with the reference (original) utterance of the same content from the same speaker. 
Each pair are rated by one rater with the following instructions: “You should not judge the content, grammar, or audio quality of the sentences; instead, just focus on the similarity of the speakers to one another”. The similarity of each pair of the sentences is rated on a scale from 1 to 5 (1: very unsimilar, 2: unsimilar, 3: slightly similar, 4: similar, 5: almost the same). 

\begin{table}[th]
  \caption{Speaker similarity Mean Opinion Score (MOS)}
  \label{tab:speaker similarity}
  \centering
  \begin{tabular}{l  c}
        \toprule
        \textbf{Method} & \textbf{MOS} \\
        \midrule 
        Baseline\cite{jia2018transfer}  & 2.69\\
        Ours & \textbf{3.41}\\
        \bottomrule
    \end{tabular}
\end{table}

The results are shown in Table~\ref{tab:speaker similarity}. Despite the fact that the baseline method focuses on extracting timbre of the target speaker, the MOS of our method is 0.72 higher. The results demonstrate the success of our method in exploring long range correlation to extract timbre information. 


\section{Conclusion}
In this paper, we have designed a one-stage context-aware framework for zero-shot text-based speech synthesis. Our method jointly considers the three components of speech, namely language content, timbre, and prosody, by an end-to-end framework that directly synthesizes the target speech. The input information is made full use of by several transformer-based modules. Experimental results show that our method can synthesize target speech with high naturalness and similarity without the training corpus of the target speaker. 

As we know, the performance of Griffin-Lim algorithm is not quite satisfactory. In our future work, we plan to jointly train a high-quality vocoder. Besides, we plan to extend our work to zero-shot TTS scenario to synthesize longer phrases or even the whole sentence. 

\bibliographystyle{IEEEtran}

\bibliography{mybib}

\begin{thebibliography}{10}
\providecommand{\url}[1]{#1}
\csname url@samestyle\endcsname
\providecommand{\newblock}{\relax}
\providecommand{\bibinfo}[2]{#2}
\providecommand{\BIBentrySTDinterwordspacing}{\spaceskip=0pt\relax}
\providecommand{\BIBentryALTinterwordstretchfactor}{4}
\providecommand{\BIBentryALTinterwordspacing}{\spaceskip=\fontdimen2\font plus
\BIBentryALTinterwordstretchfactor\fontdimen3\font minus
  \fontdimen4\font\relax}
\providecommand{\BIBforeignlanguage}[2]{{%
\expandafter\ifx\csname l@#1\endcsname\relax
\typeout{** WARNING: IEEEtran.bst: No hyphenation pattern has been}%
\typeout{** loaded for the language `#1'. Using the pattern for}%
\typeout{** the default language instead.}%
\else
\language=\csname l@#1\endcsname
\fi
#2}}
\providecommand{\BIBdecl}{\relax}
\BIBdecl

\bibitem{voxnest.org}
\url{https://blog.voxnest.com/}.

\bibitem{jin2017voco}
Z.~Jin, G.~J. Mysore, S.~Diverdi, J.~Lu, and A.~Finkelstein, ``Voco: Text-based
  insertion and replacement in audio narration,'' \emph{ACM Transactions on
  Graphics (TOG)}, vol.~36, no.~4, pp. 1--13, 2017.

\bibitem{fried2019text}
O.~Fried, A.~Tewari, M.~Zollh{\"o}fer, A.~Finkelstein, E.~Shechtman, D.~B.
  Goldman, K.~Genova, Z.~Jin, C.~Theobalt, and M.~Agrawala, ``Text-based
  editing of talking-head video,'' \emph{ACM Transactions on Graphics (TOG)},
  vol.~38, no.~4, pp. 1--14, 2019.

\bibitem{discript.org}
\url{https://www.descript.com/}.

\bibitem{jin2016cute}
Z.~Jin, A.~Finkelstein, S.~DiVerdi, J.~Lu, and G.~J. Mysore, ``Cute: A
  concatenative method for voice conversion using exemplar-based unit
  selection,'' in \emph{2016 IEEE International Conference on Acoustics, Speech
  and Signal Processing (ICASSP)}.\hskip 1em plus 0.5em minus 0.4em\relax IEEE,
  2016, pp. 5660--5664.

\bibitem{su2020acoustic}
J.~Su, Z.~Jin, and A.~Finkelstein, ``Acoustic matching by embedding impulse
  responses,'' in \emph{ICASSP 2020-2020 IEEE International Conference on
  Acoustics, Speech and Signal Processing (ICASSP)}.\hskip 1em plus 0.5em minus
  0.4em\relax IEEE, 2020, pp. 426--430.

\bibitem{morrison2021context}
M.~Morrison, L.~Rencker, Z.~Jin, N.~J. Bryan, J.-P. Caceres, and B.~Pardo,
  ``Context-aware prosody correction for text-based speech editing,''
  \emph{arXiv preprint arXiv:2102.08328}, 2021.

\bibitem{jia2018transfer}
Y.~Jia, Y.~Zhang, R.~J. Weiss, Q.~Wang, J.~Shen, F.~Ren, Z.~Chen, P.~Nguyen,
  R.~Pang, I.~L. Moreno \emph{et~al.}, ``Transfer learning from speaker
  verification to multispeaker text-to-speech synthesis,'' \emph{arXiv preprint
  arXiv:1806.04558}, 2018.

\bibitem{shen2018natural}
J.~Shen, R.~Pang, R.~J. Weiss, M.~Schuster, N.~Jaitly, Z.~Yang, Z.~Chen,
  Y.~Zhang, Y.~Wang, R.~Skerrv-Ryan \emph{et~al.}, ``Natural tts synthesis by
  conditioning wavenet on mel spectrogram predictions,'' in \emph{2018 IEEE
  International Conference on Acoustics, Speech and Signal Processing
  (ICASSP)}.\hskip 1em plus 0.5em minus 0.4em\relax IEEE, 2018, pp. 4779--4783.

\bibitem{li2019neural}
N.~Li, S.~Liu, Y.~Liu, S.~Zhao, and M.~Liu, ``Neural speech synthesis with
  transformer network,'' in \emph{Proceedings of the AAAI Conference on
  Artificial Intelligence}, vol.~33, no.~01, 2019, pp. 6706--6713.

\bibitem{ren2019fastspeech}
Y.~Ren, Y.~Ruan, X.~Tan, T.~Qin, S.~Zhao, Z.~Zhao, and T.-Y. Liu, ``Fastspeech:
  Fast, robust and controllable text to speech,'' \emph{arXiv preprint
  arXiv:1905.09263}, 2019.

\bibitem{ren2020fastspeech}
Y.~Ren, C.~Hu, T.~Qin, S.~Zhao, Z.~Zhao, and T.-Y. Liu, ``Fastspeech 2: Fast
  and high-quality end-to-end text-to-speech,'' \emph{arXiv preprint
  arXiv:2006.04558}, 2020.

\bibitem{qian2020unsupervised}
K.~Qian, Y.~Zhang, S.~Chang, M.~Hasegawa-Johnson, and D.~Cox, ``Unsupervised
  speech decomposition via triple information bottleneck,'' in
  \emph{International Conference on Machine Learning}.\hskip 1em plus 0.5em
  minus 0.4em\relax PMLR, 2020, pp. 7836--7846.

\bibitem{griffin1984signal}
D.~Griffin and J.~Lim, ``Signal estimation from modified short-time fourier
  transform,'' \emph{IEEE Transactions on acoustics, speech, and signal
  processing}, vol.~32, no.~2, pp. 236--243, 1984.

\bibitem{vaswani2017attention}
A.~Vaswani, N.~Shazeer, N.~Parmar, J.~Uszkoreit, L.~Jones, A.~N. Gomez,
  L.~Kaiser, and I.~Polosukhin, ``Attention is all you need,'' \emph{arXiv
  preprint arXiv:1706.03762}, 2017.

\bibitem{mcauliffe2017montreal}
M.~McAuliffe, M.~Socolof, S.~Mihuc, M.~Wagner, and M.~Sonderegger, ``Montreal
  forced aligner: Trainable text-speech alignment using kaldi.'' in
  \emph{Interspeech}, vol. 2017, 2017, pp. 498--502.

\bibitem{zen2019libritts}
H.~Zen, V.~Dang, R.~Clark, Y.~Zhang, R.~J. Weiss, Y.~Jia, Z.~Chen, and Y.~Wu,
  ``Libritts: A corpus derived from librispeech for text-to-speech,''
  \emph{arXiv preprint arXiv:1904.02882}, 2019.

\bibitem{kingma2014adam}
D.~P. Kingma and J.~Ba, ``Adam: A method for stochastic optimization,''
  \emph{arXiv preprint arXiv:1412.6980}, 2014.

\bibitem{wang2017tacotron}
Y.~Wang, R.~Skerry-Ryan, D.~Stanton, Y.~Wu, R.~J. Weiss, N.~Jaitly, Z.~Yang,
  Y.~Xiao, Z.~Chen, S.~Bengio \emph{et~al.}, ``Tacotron: Towards end-to-end
  speech synthesis,'' \emph{arXiv preprint arXiv:1703.10135}, 2017.

\end{thebibliography}


\end{document}